\documentclass[12pt,a4paper,twoside]{article}
\usepackage{epsfig}
\usepackage{baltlat2}
\usepackage{wrapfig}
\usepackage{dcolumn}
\begin{document}
\ \
\vspace{0.5mm}
\setcounter{page}{1}
\vspace{6mm}
\titlehead{Baltic Astronomy, vol.15, p.521--530.}
\titleb{THE OPTICAL SPECTRUM OF R CORONAE BOREALIS CLOSE TO 2003 DECLINE}

\begin{authorl}
\authorb{T\~onu Kipper}{1} and
\authorb{Valentina G. Klochkova}{2} 
\end{authorl}

\begin{addressl}
\addressb{1}{Tartu Observatory, T\~oravere, 61602, Estonia; tk@aai.ee} 

\addressb{2}{Special Astrophysical Observatory RAS, Nizhnij Arkhyz, 369167, Russia; valenta@sao.ru} 
\end{addressl}

\submitb{Received ..., 2006}
\begin{abstract}
Two sets of high-resolution spectra of R\,CrB obtained during the 2003
light decline are described. The first set was obtained on the descending
branch of the light curve when $V\approx12.0$ and the second one in the
recovery phase with $V\approx7.5$. The usual sharp and broad emissions are
described and the lines radial velocities measured. C$_2$ Swan system
(0,0) band was found to be in emission for the first set. The other C$_2$
bands were in absorption. Few CN red system (5,1) band rotational lines
and low excitation Fe\,I lines were in absorption. A table with measured
radial velocities of various spectral features is presented.
\end{abstract}
\begin{keywords}
stars: atmospheres -- stars:
 individual: R\,CrB
\end{keywords}
\resthead{R\,CrB 2003 minimum}{T.\,Kipper, V.G.\,Klochkova }

\sectionb{1}{INTRODUCTION}

R\,CrB is the prototype of peculiar supergiant stars with fast and deep
dimmings by several magnitudes at unpredictable times. These dimmings last
several weeks or months. The atmospheres of R\,CrB type stars are
extremely hydrogen deficient and carbon rich. The light minima are
believed to be caused by the formation of obscuring clouds of carbon soot
(O'Keefe 1939). About 30 stars of the class are known and the prototype is
one of the most studied since its discovery 210 years ago. Due to
unpredictability of the dimmings the spectral observations during the
light minima are sparse. However, these observations could provide novel
information about the outer stellar atmosphere and circumstellar region as
the photosphere is almost completely obscured for several weeks before the
soot cloud disperses. 

In this note we report on spectra obtained during the 2003 light minimum
of R\,CrB. {\it Clayton (1996) gave a general review of R\,CrB
spectra in and out of declines.  Rao et al. (1999) studied the optical
spectrum of R\,CrB in the very deep and prolongated minimum in 1995.}
The previous minimum of R\,CrB on 2000 was described by Kipper (2001).
The 2003 minimum was also observed by Rao et al. (2006). Our spectra were
obtained somewhat earlier and later and therefore well complement Rao et
al. data.

In Fig.\,1 the light curve of R\,CrB during the 2003 light minimum is
shown (AAVSO data) and the moments of our (long vertical lines) and Rao et
al. (2006) specroscopic observations are marked. The first set of our
spectra was obtained on February 24 with the star at $V\approx 12.0$ and
the second one on April 11 when the star has recovered to $V\approx 7.5$.
Finally, two sets of the spectra were obtained on January 12, 2004 and
April 18, 2006 when the star has completely recovered.

\begin{figure} 
\vskip2mm
\centerline{{\psfig{figure=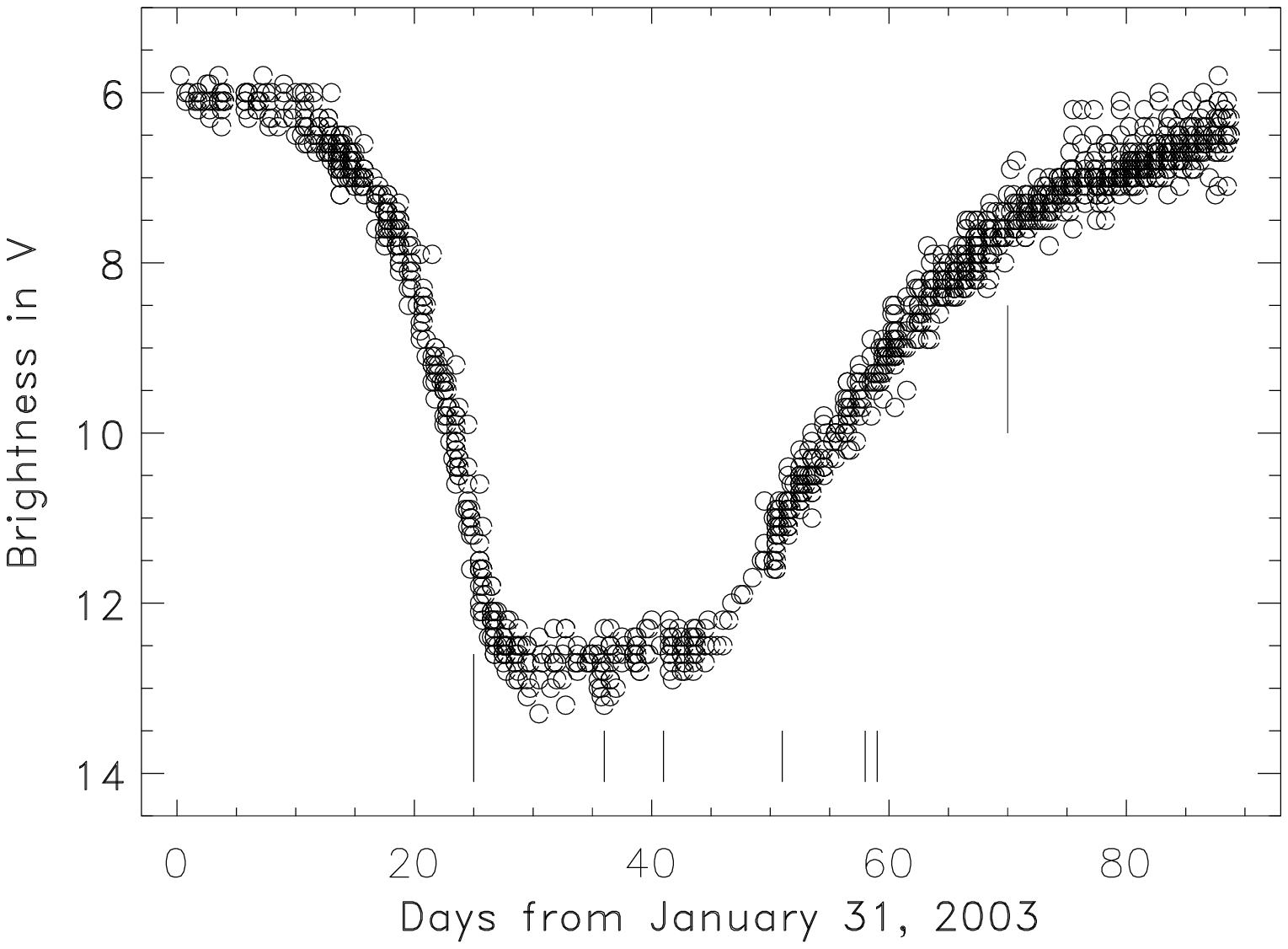,angle=0,width=120mm}}}
\vskip2mm
\captionb{1}{The light curve of R\,CrB during the 2003 decline (AAVSO).
The time is counted from February 1 2003. Longer vertical ticks indicate 
our observations and shorter ones Rao et al. (2006) spectra.
Our  observations after the full recovery are not indicated.}
\vskip2mm
\end{figure}

\sectionb{2}{OBSERVATIONS}

All our high resolution ($R\approx$\,42\,000) spectra were obtained with
the Nasmyth Echelle Spectrometer (Panchuk et al. 1999; Panchuk et al.
2002) of Russian 6 m telescope. The spectrograph was equipped with an
image slicer (Panchuk et al. 2003). As a detector a CCD camera with
$2052\times2052$ pixels produced by the Copenhagen University Observatory
was used. The first set,which was observed close to the light minimum and
the last one at the maximum light cover the wavelength region
$516.0\div666.0$\,nm without gaps up to 600.0\,nm. The set observed when
the {\it star's} brightness has almost recovered is shifted toward blue
and covers $448.0\div600.0$\,nm without gaps.

\sectionb{3}{DESCRIPTION OF THE SPECTRA}

\subsectionb{3.1}{Reduction of the spectra}

The spectra were reduced using the NOAO astronomical data analysis
facility IRAF. The continuum was placed by fitting low order spline
functions through the manually indicated points in every order. The use of
image slicer results in three parallel strips of spectra in each order.
These strips are wavelength shifted. Therefore all strips were reduced
separately and then already linearized in the wavelength spectra were
coadded. We checked the accuracy of this procedure (Kipper \& Klochkova
2005) and found that the wavelengths of the terrestial lines in the
stellar spectrum were reproduced within a few 0.001\,\AA-s. After that all
spectra of the set were coadded.
 
As measured from the Th-Ar comparison spectra the resolution is
$R\approx$\,42\,800 with FWHM of comparison lines about 7\,km\,s$^{-1}$.

{\it As it was expecting for the spectrum of R\,CrB close to its
minimum light, the spectrum by February 24 contains numerous
emission sharp lines, the complex profile of Na\,I\,D lines
including broad and sharp emissions, weak H$\alpha$ emission,
strong C\,I lines, the remarkable emission of forbidden [O\,I],
the $C_2$, CN molecular bands. All these spectral features were
considered below in detail.

It should be noted here that main abovementioned spectral peculiarities
caused by circumstellar gas are also observed in spectra of selected
post-AGB stars with circumstellar envelopes. The most appropriate example
is a semiregular variable star QY\,Sge identified with the IR-source
IRAS\,20056+1834 (Rao et al. 2002). But, in contrast to R\,CrB stars,
typical spectral peculiarities in spectra of post-AGB stars are
independent on observing moment, the are permanently visible.} 

\subsectionb{3.2}{The sharp emission lines}

As during the earlier declines of R\,CrB the emission lines dominate the
spectrum and most lines could be classified as sharp lines of neutral and
singly ionized metals. By April 11, when the star was still by 1.5
magnitudes fainter than in maximum, the sharp emission lines have
dissapeared. The mean heliocentric radial velocity of these sharp lines is
$v_{rad}=21.0\pm0.5$\,km\,s$^{-1}$ and mean FWHM is about
$22.2\pm1.9$\,km\,s$^{-1}$. Therefore we confirm Rao et al. (2006) findings
that the sharp lines in 2003 are somewhat broader than in 1995-1996, when
their FWHM was about 15\,km\,s$^{-1}$, and that the sharp emission lines
were not shifted relative to the mean stellar velocity at maximum light
($+22.5$\,km\,s$^{-1}$). During the earlier dimming in 2000 the sharp
emission lines showed the blueshift around 6\,km\,s$^{-1}$ relative to
mean systemic velocity (Kipper 2001).

\begin{figure}
\vskip2mm
\centerline{{\psfig{figure=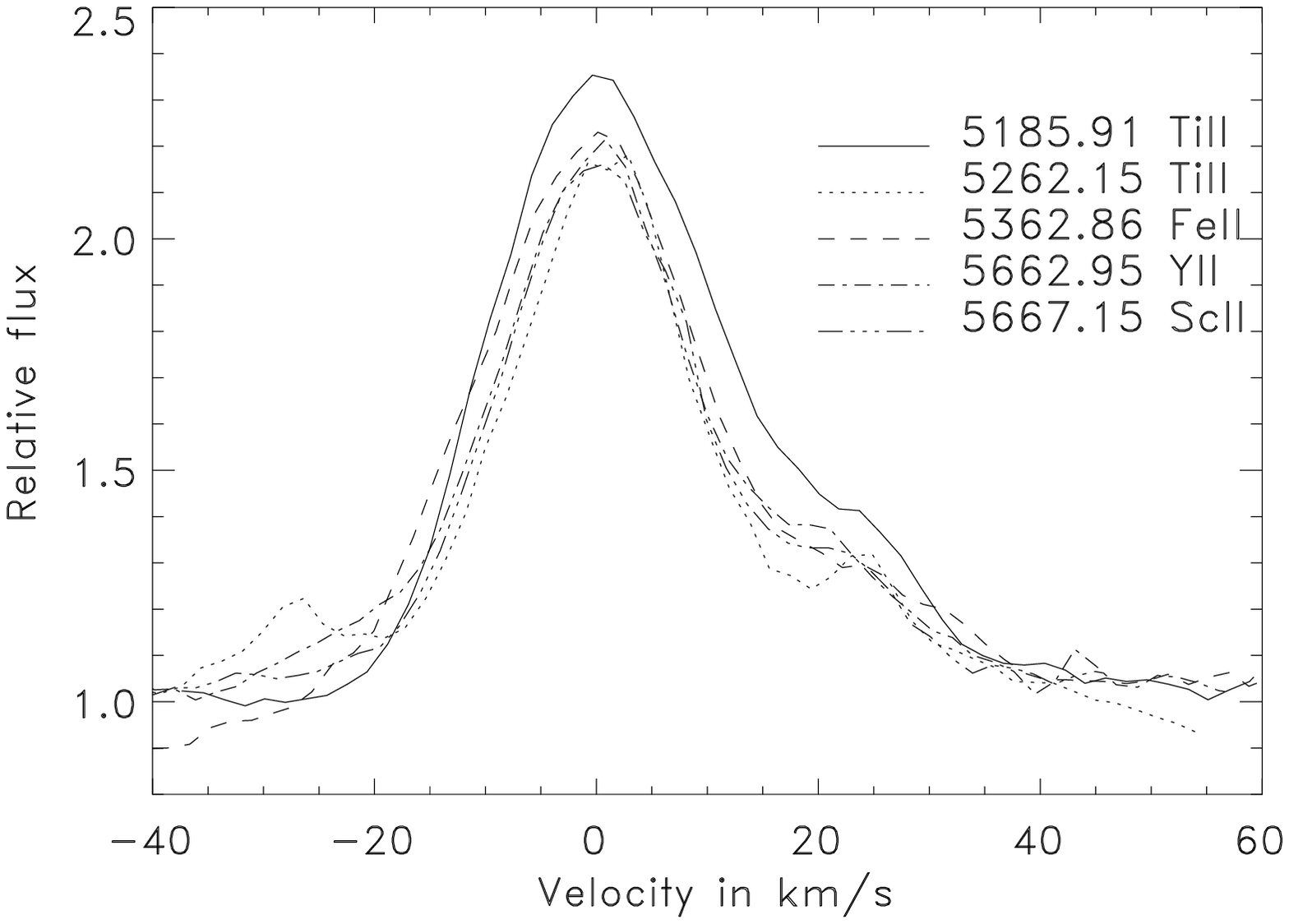,width=120mm}}}
\vskip2mm
\captionb{2}{The profiles of some emission lines in velocity scale.
The velocity scale is set for the blue components.}
\vskip2mm
\end{figure}

\begin{figure}
\vskip2mm
\centerline{{\psfig{figure=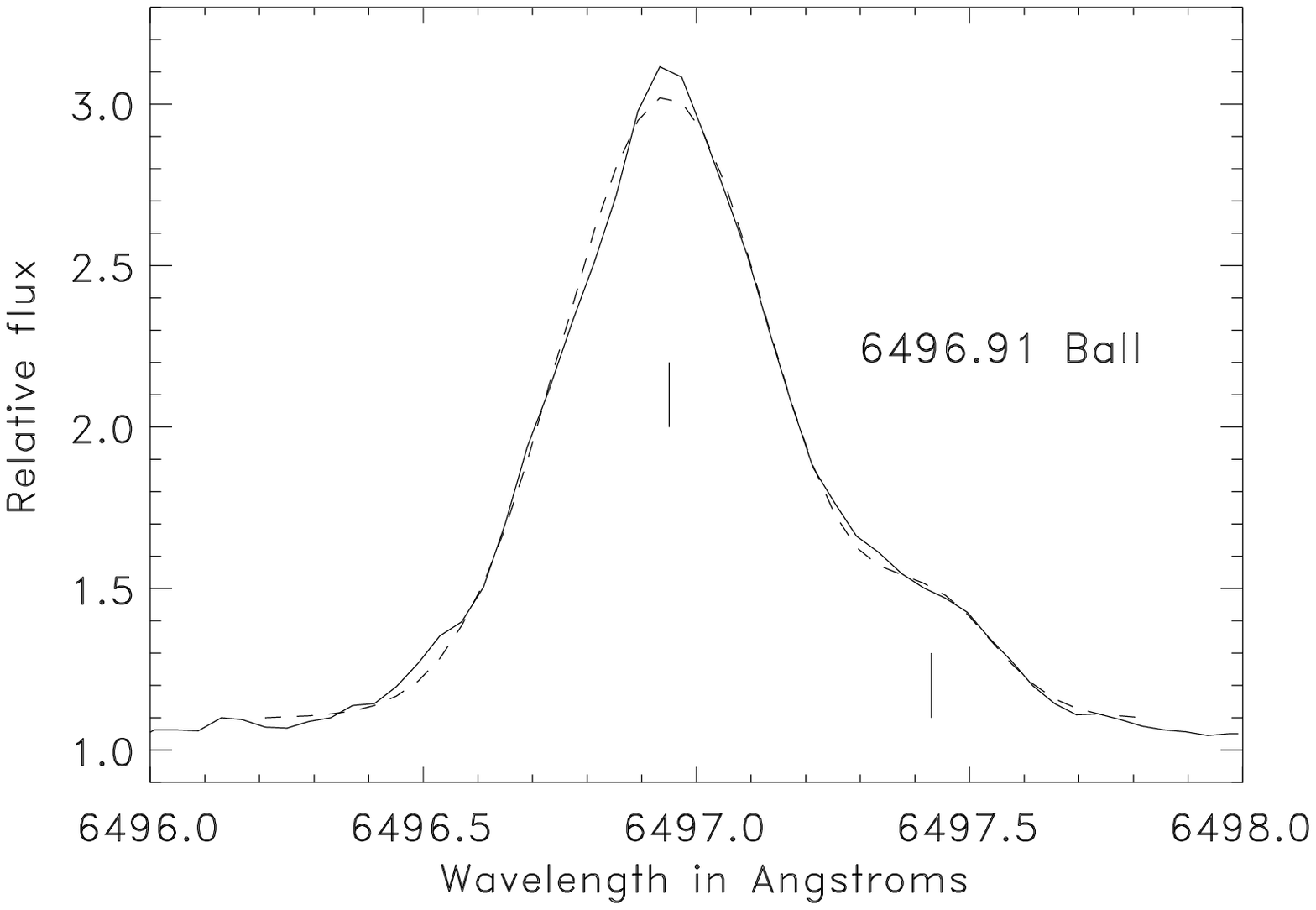,width=120mm}}}
\vskip2mm
\captionb{3}{The gaussian decomposition of the Ba\,II line at 649.691\,nm.
The positions of the components are indicated by the vertical lines.
The sum of the gaussian components is drawn by dashed line.}
\vskip2mm
\end{figure}

Most of the sharp emission lines have red components which are shifted by
about 22\,km\,s$^{-1}$ and the widths somewhat less than of the main
(Fig.\,2) components. As an example in Fig.3 the gaussian decomposition of
Ba\,II line at 649.691\,nm is presented. The main component of this line
has the heliocentric radial velocity 21.5\,km\,s$^{-1}$ and FWHM
20.8\,km\,s$^{-1}$. The red component is shifted by 22.2\,km\,s$^{-1}$ and
has FWHM=12.9\,km\,s$^{-1}$. During the 2000 light decline the sharp
emission lines had inverse P\,Cygni profile with red absorption shifted by
 $42.0\pm2.5$\,km\,s$^{-1}$ from the emission core (Kipper 2001). During 
 the 1995 minimum these absorption components were also visible (Rao
et al. 1999). Such inverse P\,Cygni components were not reported for
the 2003 decline by Rao et al. (2006). However, their Fig.\,10 shows weak
absorption components at both sides of sharp emission lines.

He\,I is presented only by weak emission line at 587.567\,nm. This line
usually shows broad emission during light declines. In the February 24
spectra the line is sharp with FWHM=24.5\,km\,s$^{-1}$ and shows red
emission component similarily to other sharp lines.

In the observed spectral {\it region} C\,I is presented only with two
emission lines at 569.313 and 667.553 nm with heliocentric velocities of
15.8 and 16.8\,km\,s$^{-1}$. This means that the C\,I lines are blueward
shifted about 6\,km\,s$^{-1}$ relative to other sharp lines.

From forbidden lines only the [O\,I] lines at 630.031 and 557.734\,nm were
present on February 24 spectra. The first [O\,I] line is blended with
Sc\,II line at 630.074\,nm. The gaussian decomposition of that blend is
depicted in Fig.\,4. The heliocentric velocity of the [O\,I] line is
21.0\,km\,s$^{-1}$. The Sc\,II line also has a red emission component
shifted by 20.5\,km\,s$^{-1}$.

We were not able to identify nonmolecular absorption lines in February 24
spectra, with possible exception of few nearly zero-exci\-ta\-tion
($\varepsilon_i=0.11\div0.12$) Fe\,I lines with velocities about
14\,km\,s$^{-1}$. The identification of these lines is uncertain. In this
respect the 2003 decline differs from the 2000 minimum when we found rich
absorption spectrum. The strongest lines belonged to C\,I but the lines of
Fe\,I, Fe\,II, Cr\,I, O\,I, Mg\,I, Na\,I and Si\,II were also present
(Kipper 2001).

\begin{figure}
\vskip2mm
\centerline{{\psfig{figure=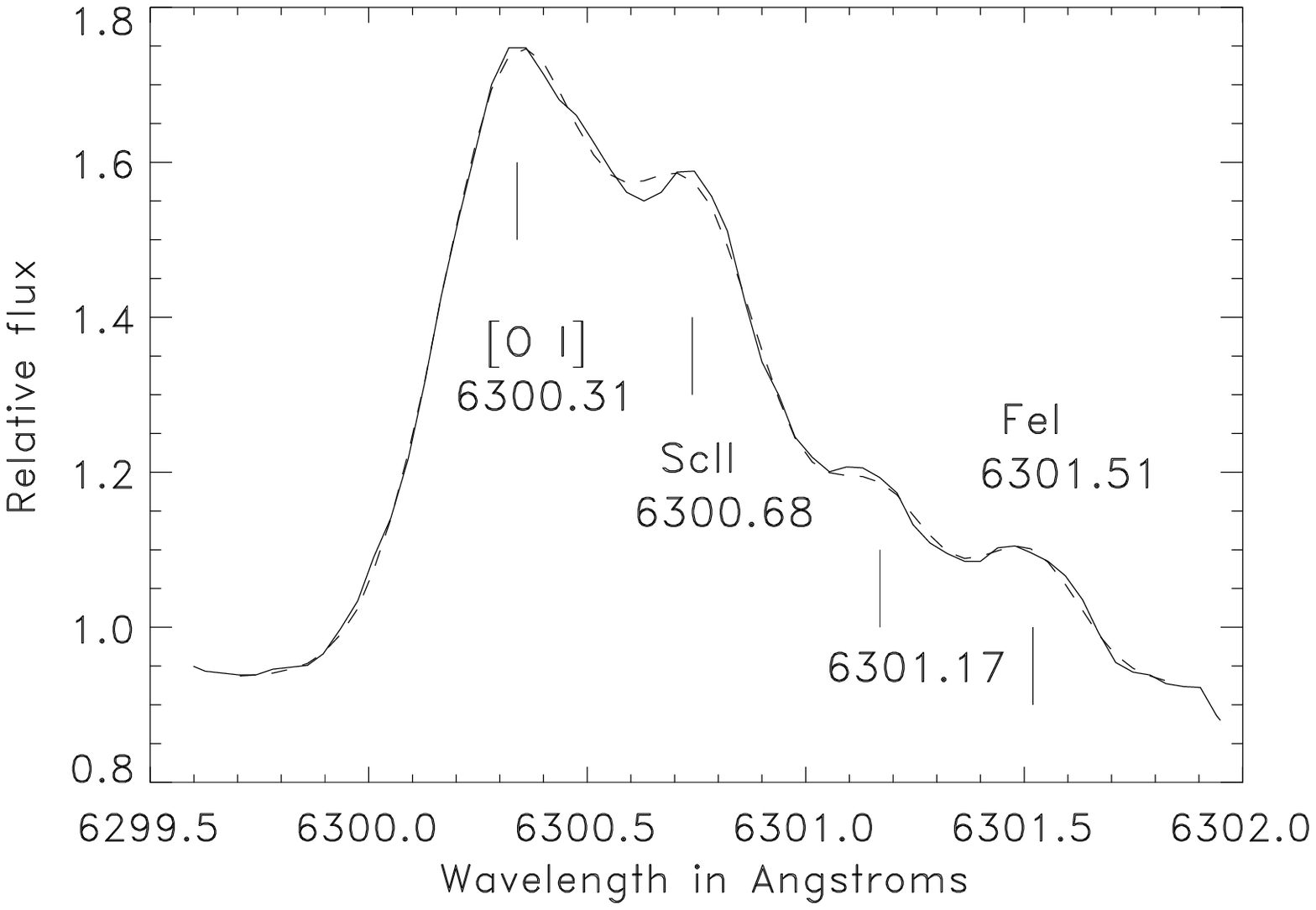,width=120mm}}}
\vskip2mm
\captionb{4}{The gaussian decomposition of the blend of [O\,I] and
Sc\,II lines at 630.05\,nm. The solar wavelengths are indicated for the
identified lines. The measured value is indicated for the Sc\,II red
component.}
\vskip2mm
\end{figure}

\subsectionb{3.3} {The broad emission lines}

From the category of broad emission lines only the Na\,I\,D lines were
visible in February 24, 2003 spectra (Fig.\,5). As during the earlier
declines the Na\,I {\it doublet} consists of broad and sharp emission
components. Also, the sharp interstellar (IS) lines of the {\it doublet}
were visible. The mean over three observing epochs heliocentric velocity
of IS lines is $22.1\pm1.1$\,km\,s$^{-1}$. The stated error could be
considered as the possible error of all our radial velocity mearurements.

\begin{figure} 
\vskip2mm
\centerline{{\psfig{figure=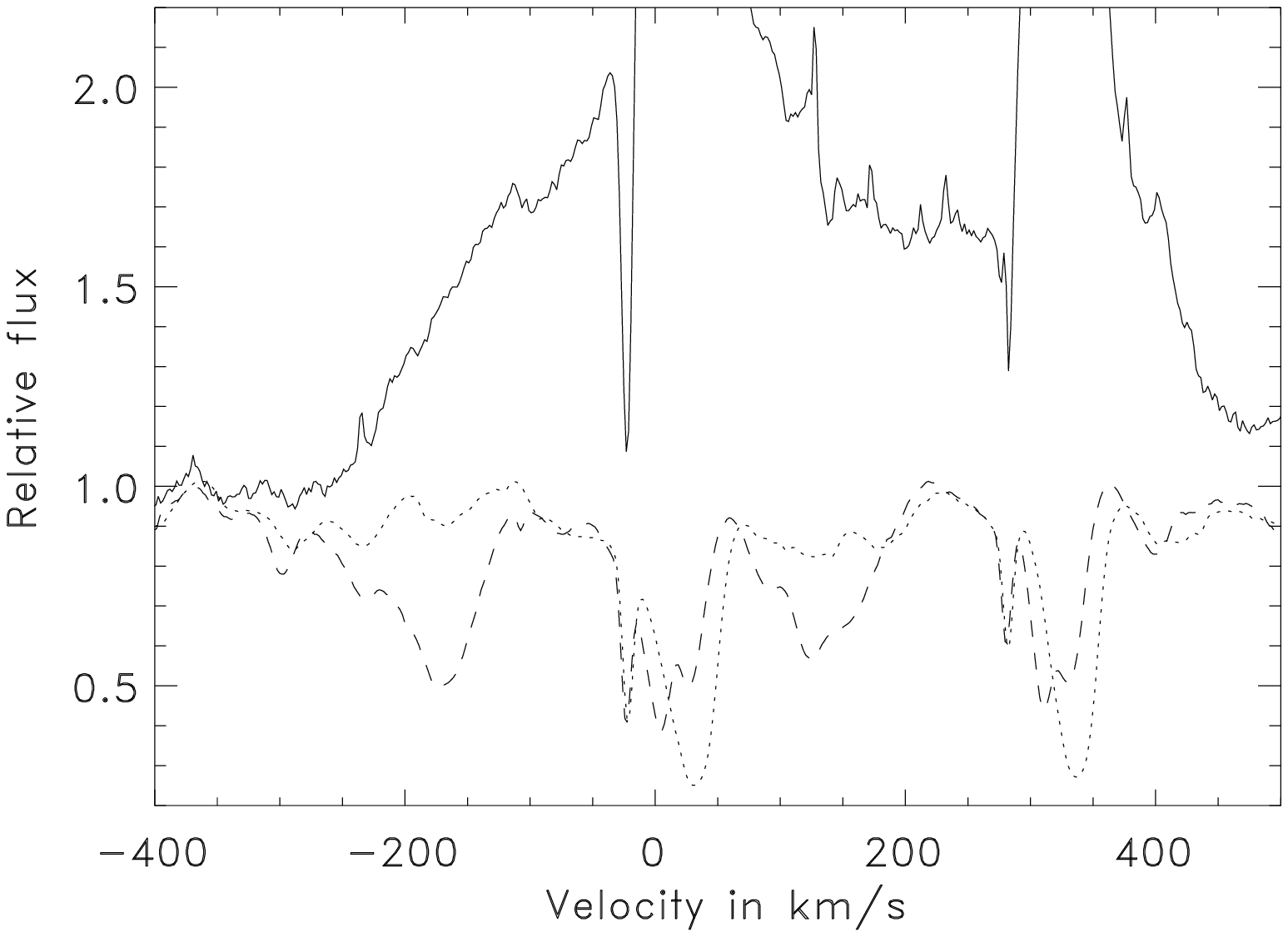,angle=0,width=120mm}}}
\vskip2mm
\captionb{5}{Observed spectra of R\,CrB near the 
Na\,I\,D {\it doublet}. The velocity
scale is set for the D$_2$ line. Full line -- specrum for February 24,
dashed line -- the one for April 11, 2003, and dotted line -- spectrum in
light maximum (January 12, 2004).}
\vskip2mm
\end{figure}

The sharp emissions of Na\,D lines are slightly {\it blueshifted} 
($\approx3$\,km\,s$^{-1}$) relative to other sharp emission lines and
FWHM=21.4\,km\,s$^{-1}$. The broad component of D$_2$ line extends to
$-266$\,km\,s$^{-1}$. Superimposed on this broad wing is an absorption
component at $-99$\,km\,s$^{-1}$. The absorption nearly at the same
velocity was found also by Rao et al. (2006).

In the April spectra two strong blueshifted absorption components of D$_2$
line are visible at heliocentric velocities $-230$ and
$-172$\,km\,s$^{-1}$. The absorption profile in D$_1$ line is more
complicated. The most blueshifted components are at the velocities of
$-218$ and $-178$\,km\,s$^{-1}$. Such high-velocity gas has also been
reported for other declines of R\,CrB stars after minimum light (Rao et
al. 2004).

The photospheric absorption of Na\,I consists of two components at
heliocentric radial velocities of 4.4 and 31.5\,km\,s$^{-1}$ (D$_2$), and
6.5 and 29.9\,km\,s$^{-1}$ (D$_1$). The separation of these components
could be caused by sharp emission in the line core.

The H$\alpha$ line shows weak two-peaked emission in the February spectra.
 This emission was blueshifted relative to the photospheric absorption. In the
recovery phase the line is in absorption as it is in the light maximum 
(Fig.\,6).

\begin{figure} 
\vskip2mm
\centerline{{\psfig{figure=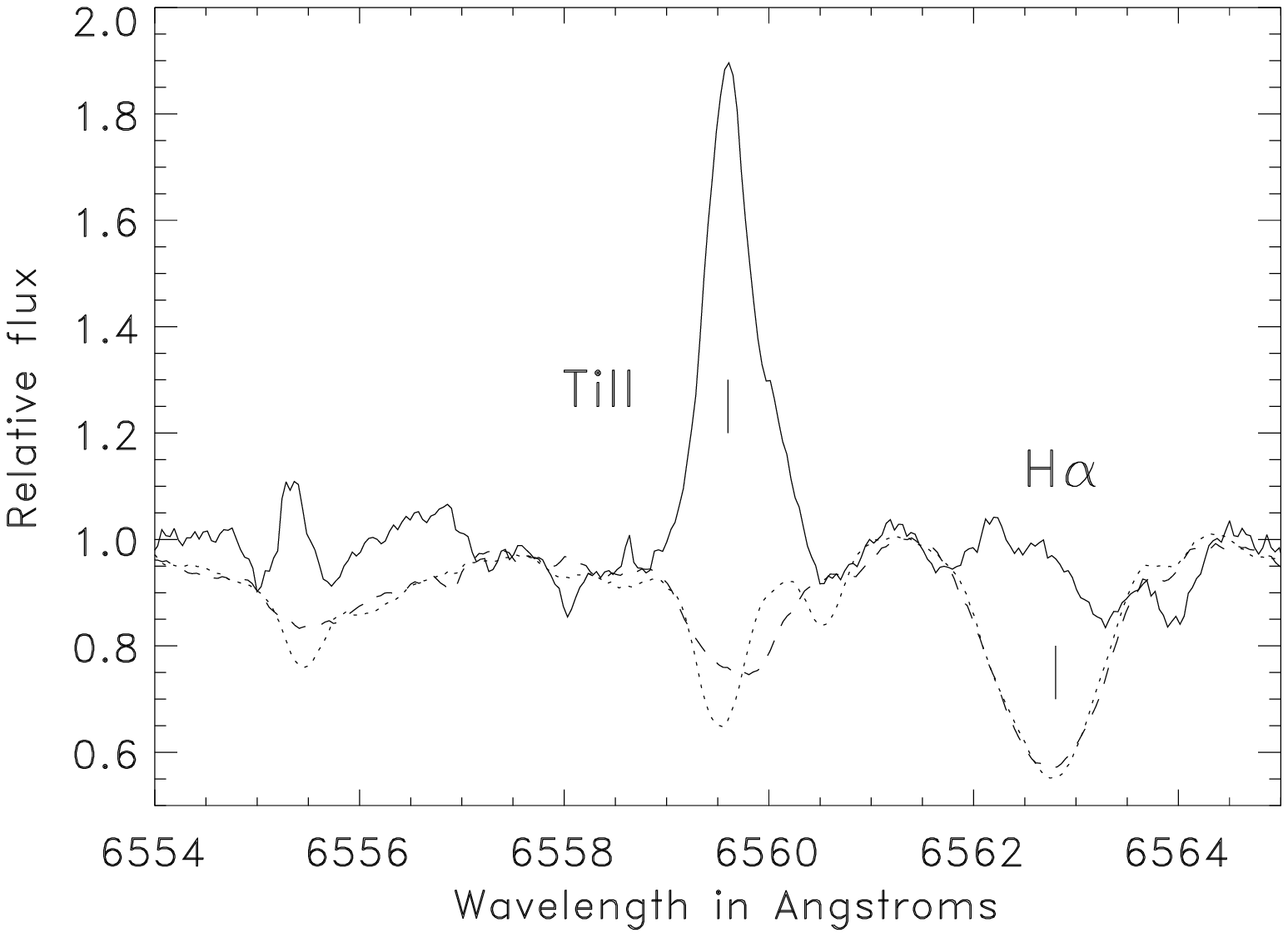,angle=0,width=120mm}}}
\vskip2mm
\captionb{6}{The spectrum of R\,CrB near the H$\alpha$ line.
Full line -- spectrum on February 24, 2003, dotted line -- spectrum
on April 11, 2003, and dashed line -- spectrum on April 18, 2006.}
\vskip2mm
\end{figure}

\subsectionb{3.4} {The molecular lines}

Rao et al. (2006) reported that their spectra of R\,CrB during the late
phases of the 2003 dimming show C$_2$ Swan bands (1,2), (0,2), and (1,4)
in emission. Our spectra for February 24, 2003 show only the C$_2$ Swan
system (0,0) band in emission (Fig.\,7). The lines of this band are
broadened so that the {\it rotational} structure is not visible. The other
observed bands (1,2), (0,1), (0,2), and (1,3) were in absorption
(Fig.\,8). The absorption lines are sharp with
FWHM$\approx22$\,km\,s$^{-1}$. We were able to determine velocities from
the rotational lines of (0,1) and (0,2) bands $16.0\pm1.0$ and
$15.4\pm2.6$\,km\,s$^{-1}$ correspondingly. Some absorption lines of CN
red system (5,1) band were identified with the velocity of
$15.0\pm2.3$\,km\,s$^{-1}$. This means that the molecular absorption lines
were blueshifted relative the systemic velocity by 7\,km\,s$^{-1}$.

\begin{figure} 
\vskip2mm
\centerline{{\psfig{figure=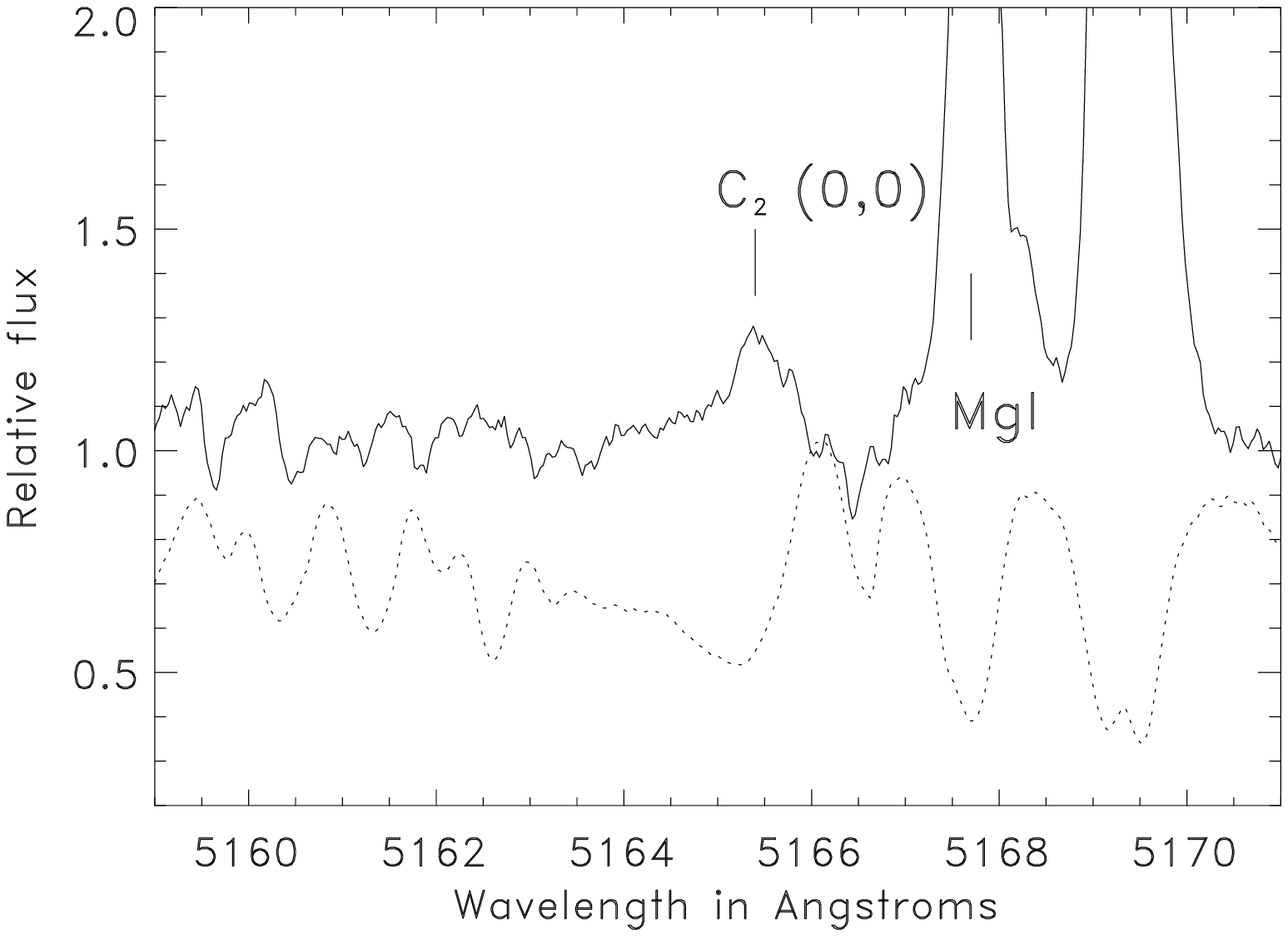,angle=0,width=120mm}}}
\vskip2mm
\captionb{7}{The portion of the spectrum of R\,CrB close to the C$_2$ Swan
system (0,0) band head. Full line -- spectrum at February 24, dotted
line -- at April 11, 2003. The emission lines of Mg\,I and the emission
blend of Fe\,I and Fe\,II are also visible.}
\vskip2mm
\end{figure}

\begin{figure}
\vskip2mm
\centerline{{\psfig{figure=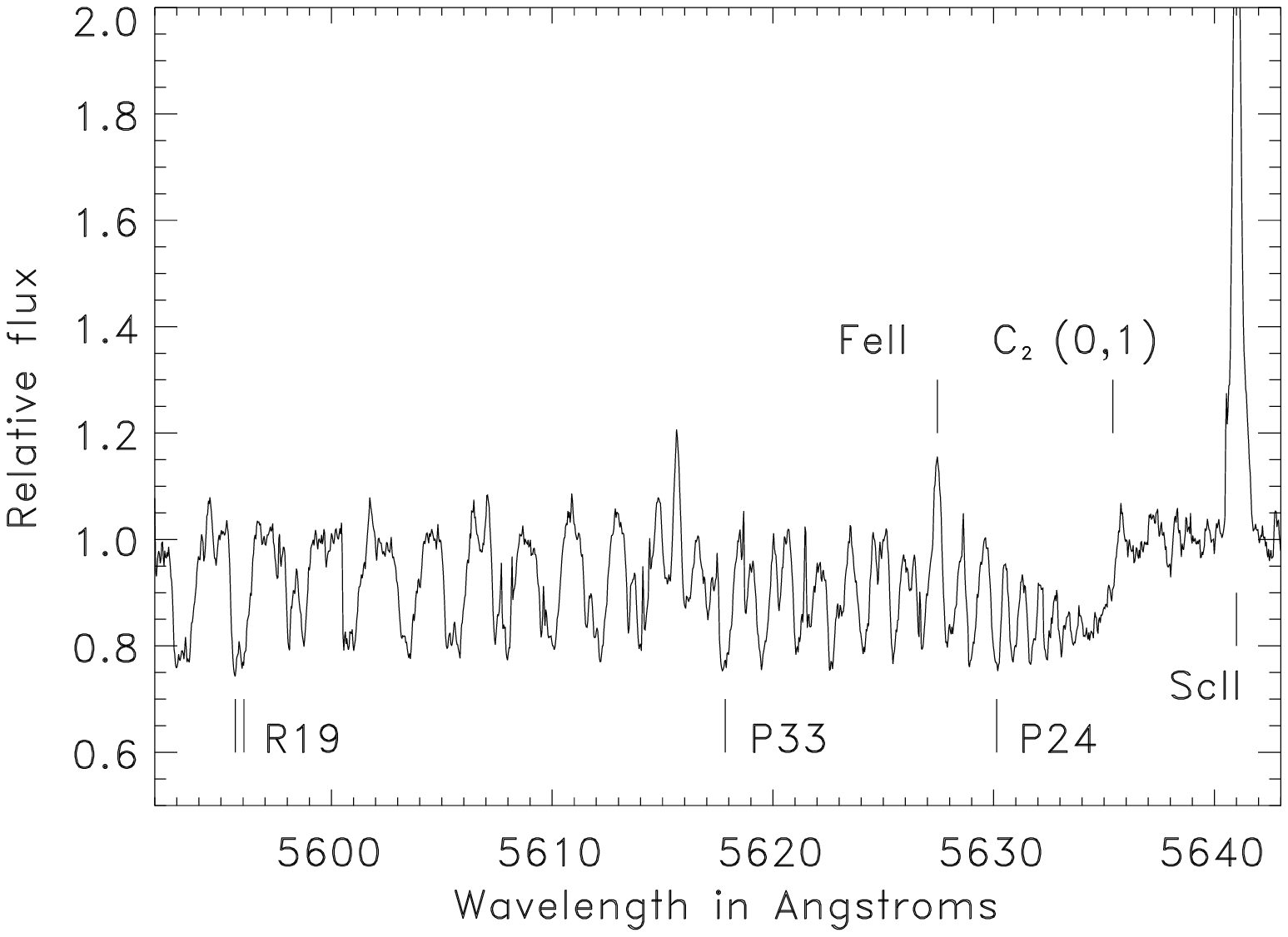,angle=0,width=120mm}}}
\vskip2mm
\captionb{8}{The portion of the spectrum of R\,CrB close to the S$_2$ 
 Swan system
(0,1) band head
on February 24, 2003. The
 emission lines of Fe\,II and Sc\,II are indicated.
The rotational lines of C$_2$ are sharp and well resolved.} 
\vskip2mm
\end{figure}

\vbox{  
\begin{center}
\vbox{\footnotesize
\begin{tabular}{lccc}
\multicolumn{4}{c}{\parbox{100mm}{\baselineskip=9pt
{\smallbf\ \ Table 1.}{\small\ Heliocentric radial velocities
of various spectral lines in the
spectrum of R\,CrB during the 2003 light decline and in maximum
(January, 12 2004).}}}\\ [+5pt]
\tablerule
Line& Feb. 24 & April 11 & Jan. 12, 2004. \\
\tablerule
Na\,I D IS: & & & \\
D$_2$ & $-23.2$ & $-23.6$ & $-21.9$ \\
D$_1$ & $-21.2$ & $-21.0$ & $-21.9$ \\ [+5pt]
Na\,I D sharp emission: & & & \\
D$_2$ & +19.1 & & \\
D$_1$ & +19.6 & & \\ [+5pt]
Na\,I D broad emission: & & & \\
Blue limit of D$_2$ & $-266$ & & \\
Abs. on blue wing & $-99$ & & \\ [+5pt]
Na\,I photosph. absorption: & & & \\
D$_2$ &  & +4.4 & +30.5 \\
      &  & +31.9 &      \\
D$_1$ &  & +6.5 & +33.0 \\
      &  & +29.9 &      \\ [+5pt]
Na\,I D blue absorption: & & & \\
D$_2$ & & $-230$ & \\
      & & $-172$ & \\
D$_1$ & & $-218$ & \\
      & & $-178$ & \\
      & & $-145$ & \\ [+5pt]
Sharp emissions: & & & \\ 
Main component & +21.0 & & \\
Red component  & +43   & & \\
C\,I emission & +16.3 & & \\
$[$O\,I$]$ emission & +21.0 & & \\ [+5pt]
C$_2$ rot. lines absorption & +15.7 & & \\
CN rot.lines absorption & +15.0 & & \\
Fe\,I absorption & +14 & & \\
\tablerule
\end{tabular}
}
\end{center}
}

\sectionb{4}{CONCLUSION}

We have presented the spectral observations of R\,CrB during the 2003 minimum
and compared them with earlier light declines.

The sharp emission line spectrum is clearly universal during all light
drops. Usually the sharp emission lines are blueshifted by about
10\,km\,s$^{-1}$. In 2003 these lines were almost not shifted. At the same
time the lines widths are larger than usually. Already the usual widths
around 15\,km\,s$^{-1}$ are much larger than if caused by thermal
broadening. The small or absent blueshift together with large linewidth
could indicate that the line-forming region is expanding roughly with
$v_{exp}\approx10$\,km\,s$^{-1}$.

In 2003 the sharp lines did not show inverse P\,Cygni profiles which were
prominent in 2000. Instead, the lines showed red emission components.

The quite universal are also the broad and blueshifted absorptions of
Na\,I D lines developing in the recovery phase.

On the descending part of the light curve C$_2$ Swan bands were in
absorption except the (0,0) band. Later, during minimum light these
bands were in emission. At the maximum light the C$_2$ Swan system
bands are weakly in absorption.

\vskip5mm

ACKNOWLEDGEMENTS.\ This research was supported by the Estonian Science
Foundation grant nr.~6810 (T.K.). V.G.K. acknowledges the support from the
programs of Russian Academy of Sciences ``Observational manifestations of
evolution of chemical composition of stars and Galaxy'' and ``Extended
objects in Universe''. V.G.K. also acknowledges the support by Award
No.\,RUP1--2687--NA--05 of the U.S. Civilian Research \& Development
Foundation (CRDF). We acknowledge with thanks the observations from the
AAVSO International Database.

\goodbreak

\References{}
\refb
{\it Clayton G.C. 1996, PASP, 108, 225}
\refb
Kipper T. 2001, IBVS, 5063
\refb
Kipper T., Klochkova V.G. 2005, Baltic Astronomy, 14, 215
\refb
O'Keefe J.A. 1939, ApJ, 90, 294
\refb
Panchuk V.E., Klochkova V.G., Naidenov I.D. 1999, Preprint
Spec. AO, No 135
\refb
Panchuk V.E., Piskunov N.E., Klochkova V.G., et al. 2002, Preprint
Spec. AO, No 169
\refb
Panchuk V.E., Yushkin M.V., Najdenov I.D. 2003, Preprint Spec. AO, No 179
\refb
Rao N.K., Lambert D.L., Adams M.T., et al. 1999, MNRAS, 310, 717
\refb
{\it Rao N.K., Goswami A., Lambert D.L. 2002, MNRAS, 334, 129}
\refb
Rao N.K., Reddy B.E., Lambert D.L. 2004, MNRAS, 355, 855
\refb
Rao N.K., Lambert D.L., Shetrone M.D. 2006, MNRAS, 370, 941
\end{document}